\documentclass{article}

\usepackage{PRIMEarxiv}

\usepackage[utf8]{inputenc} 
\usepackage[T1]{fontenc}    
\usepackage{hyperref}       
\usepackage{url}            
\usepackage{booktabs}       
\usepackage{amsfonts}       
\usepackage{nicefrac}       
\usepackage{microtype}      
\usepackage{lipsum}
\usepackage{fancyhdr}       
\usepackage{graphicx}       
\usepackage{caption}
\usepackage{tabulary} 
\usepackage[flushleft]{threeparttable}
\usepackage{multirow}
\usepackage{adjustbox}
\graphicspath{{media/}}     
\usepackage{subcaption}
\usepackage{amsmath,amsfonts,bm}
\usepackage[toc,page]{appendix}
\usepackage[colorinlistoftodos,prependcaption,textsize=tiny]{todonotes}
\usepackage{float}  
\usepackage{algorithm} 
\usepackage{algpseudocode} 
\usepackage{makecell}
\usepackage[normalem]{ulem}
\usepackage{hyperref}

\title{HelixFold-Multimer: Elevating Protein Complex Structure Prediction to New Heights
}

\author{
  Xiaomin Fang, Jie Gao, Jing Hu, Lihang Liu, Yang Xue, Xiaonan Zhang, Kunrui Zhu \\
  PaddleHelix Team\thanks{Corresponding author: Xiaomin Fang (fangxiaomin01@baidu.com). The authors are listed in alphabetical order according to their last names.} \\
  Baidu Inc. \\
  \texttt{\url{https://paddlehelix.baidu.com/}} \\
}

\begin{document}
\maketitle

\begin{abstract}
While monomer protein structure prediction tools boast impressive accuracy, the prediction of protein complex structures remains a daunting challenge in the field. This challenge is particularly pronounced in scenarios involving complexes with protein chains from different species, such as antigen-antibody interactions, where accuracy often falls short. Limited by the accuracy of complex prediction, tasks based on precise protein-protein interaction analysis also face obstacles. In this report, we highlight the ongoing advancements of our protein complex structure prediction model, HelixFold-Multimer, underscoring its enhanced performance. HelixFold-Multimer provides precise predictions for diverse protein complex structures, especially in therapeutic protein interactions. Notably, HelixFold-Multimer achieves remarkable success in antigen-antibody and peptide-protein structure prediction, greatly surpassing AlphaFold 3.
HelixFold-Multimer is now available for public use on the PaddleHelix platform, offering both \href{https://paddlehelix.baidu.com/app/drug/protein-complex/forecast}{a general version} and \href{https://paddlehelix.baidu.com/app/drug/KYKT/forecast}{an antigen-antibody version}. Researchers can conveniently access and utilize this service for their development needs.


\end{abstract}


\section{Introduction}
Understanding protein structures is vital for decoding their diverse functions, aiding drug discovery, and revealing evolutionary relationships. These structures elucidate enzymatic activity, signal transduction, and regulatory mechanisms. Accurately predicting protein structures enables protein engineering, driving biotechnological innovations for biomedical and industrial applications. Though current tools information.\cite{fang2023method,jumper2021highly,doi:10.1126/science.abj8754} can predict protein monomer structures with considerable accuracy, accurately predicting the structures of protein complexes containing multiple chains remains a formidable challenge. This difficulty arises from the complexity of capturing the interactions between chains within protein complexes, which cannot be adequately addressed solely through co-evolutionary.

Computational protein docking tools, such as ZDock \cite{chen2003zdock}, HDock \cite{yan2020hdock}, ClusPro \cite{kozakov2017cluspro}, and HADDOCK \cite{de2010haddock}, offer a cost-effective alternative, utilizing various sampling techniques to explore the conformational space and employing scoring functions to evaluate and identify optimal conformations. Despite their invaluable contributions to structural bioinformatics, these tools encounter limitations such as constraints in scoring function accuracy, conformational sampling complexities, and the treatment of protein flexibility, potentially compromising the precision of their predictions.

Recent developments in end-to-end deep learning-based protein complex approaches \cite{evans2021protein,osti_10320123,doi:10.1126/science.ade2574,drake2022protein, zheng2024improving}, exemplified by AlphaFold-Multimer \cite{evans2021protein}, aim to simultaneously fold and dock proteins within a complex, directly predicting complex structures from protein sequences. This integrated Fold and Dock methodology represents a novel problem-solving approach that notably improves the precision of structure prediction for various protein complex types. While AlphaFold-Multimer has demonstrated commendable precision compared to preceding methods for predicting protein complex structures, its performance still falls short in numerous scenarios. Nonetheless, previous studies \cite{yin2022benchmarking,yin2023evaluation,bryant2022improved} highlighted the limitations of AlphaFold-Multimer, particularly evident when tackling protein complexes that lack extensive paired Multiple Sequence Alignments (MSAs). This deficiency becomes especially apparent in modeling antigen-antibody complexes, alongside other adaptive immune recognition mechanisms, and in complexes comprising chains from diverse species. Due to the insufficient accuracy in predicting the structures of these protein complexes, researchers encounter challenges in conducting thorough analyses of protein interactions. This limitation not only hinders the understanding of the protein-protein interactions but also restricts the potential for innovative applications in protein engineering and design.

We introduce HelixFold-Multimer, a novel approach that significantly enhances the accuracy and widens the application scope of protein complex structure prediction. Building upon the groundwork laid by our prior work, HelixFold \cite{wang2022helixfold} and HelixFold-Single \cite{fang2023method}, HelixFold-Multimer improves cross-chain interaction modeling by integrating domain expertise into model architecture, input features, and training strategies.
HelixFold-Multimer is designed to provide highly accurate predictions for a diverse array of protein complex structures, particularly those involved in therapeutic protein interactions. We find that HelixFold-Multimer exhibits impressive effectiveness in accurately predicting a wide range of complex protein structures, notably excelling in predicting antigen-antibody and peptide-protein interfaces. Delineating the epitope regions of the antigen allows for further improvement in the prediction accuracy of antigen-antibody structure. Furthermore, HelixFold-Multimer demonstrates exceptional performance with commonly studied protein targets in drug development, suggesting its potential to optimize the protein design process for therapeutic development.
HelixFold-Multimer currently consists of two versions: a general version designed for predicting common protein complex structures, particularly interfaces between peptides and proteins, and a specialized version developed for predicting antigen-antibody structures. Both the \href{https://paddlehelix.baidu.com/app/drug/protein-complex/forecast}{general version} and the \href{https://paddlehelix.baidu.com/app/drug/KYKT/forecast}{antigen-antibody version} are publicly available for use on the \href{https://paddlehelix.baidu.com}{PaddleHelix platform}.

\section{Results}
\subsection{Results of General Version}
We first present the results of the general version of HelixFold-Multimer. We compare its performance on heteromeric protein complexes and peptide-protein complexes.

\subsubsection{Overall Performance of General Proteins }
\begin{figure}
    \centering
    \includegraphics[width=1.0\linewidth]{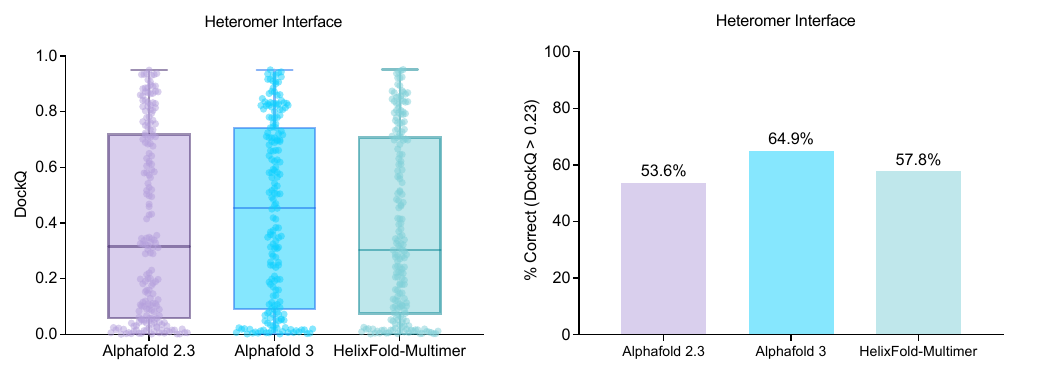} 
    \caption{Comparison between AlphaFold 2.3, AlphaFold 3 and HelixFold-Multimer for heteromeric protein complexes. The box plots on the left show DockQ score distributions, while the bar graph on the right reflects the percentage of accuracy (DockQ $>$ 0.23).}
    \label{fig:results_heter}
\end{figure}

\begin{figure}
    \centering
    \includegraphics[width=1.0\linewidth]{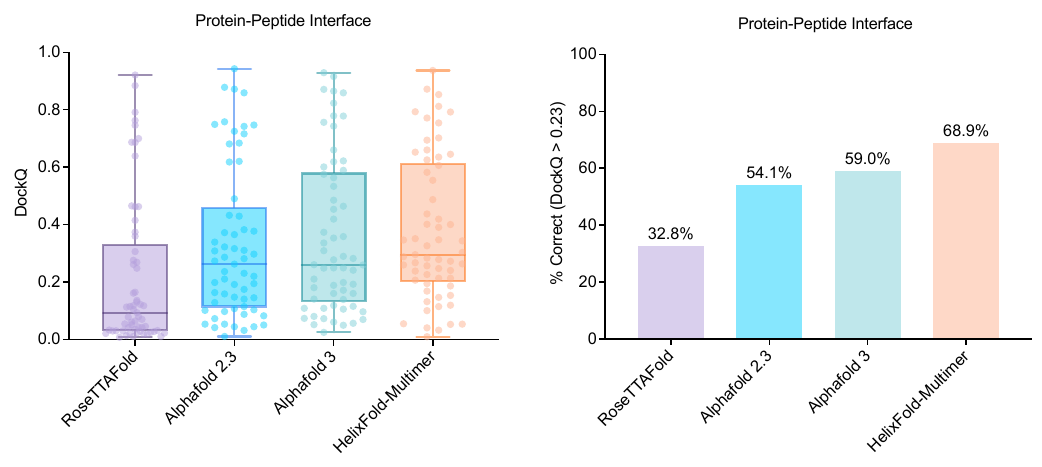} 
    \caption{Comparison of AlphaFold 2.3, AlphaFold 3 and HelixFold-Multimer in protein-peptide docking accuracy. The box plots on the left show DockQ score distributions, while the bar graph on the right reflects the percentage of accuracy (DockQ $>$ 0.23).}
    \label{fig:results_peptide}
\end{figure}

Heteromeric complexes are formed by the interaction of multiple distinct protein chains. The comparison of HelixFold-Multimer with AlphaFold 2.3 \cite{evans2021protein}, AlphaFold 3 \cite{abramson2024accurate} and RosettaFold \cite{baek2021accurate} on heteromeric protein complexes is summarized in Figure \ref{fig:results_heter}. The mean DockQ score of HelixFold-Multimer (0.378) is higher than that of AlphaFold 2.3 (0.316), but lower than AlphaFold 2.3 (0.438). When considering the percentage of the correct samples with DockQ > 0.23, AlphaFold 3 shows a leading performance with 64.9\% of its predictions above 0.23, demonstrating its robustness in protein structure prediction compared to AlphaFold 2.3 and HelixFold-Multimer. HelixFold-Multimer, while slightly behind AlphaFold 3, still outperforms AlphaFold 2.3,
This result indicates that HelixFold Multimer offers predictive performance on par with AlphaFold in modeling heteromeric protein complexes.

Peptides, being shorter chains of amino acids compared to proteins, can display a wide range of flexible structures. The challenge in predicting the precise structures of protein-peptide complexes lies in accurately modeling the flexibility of the peptides. The results of protein-peptide interfaces is shown in Figure \ref{fig:results_peptide}. HelixFold-Multimer achieved a median DockQ score of 0.295, marking a pronounced improvement compared to the median DockQ score of 0.260 obtained by AlphaFold 3 and 0.093  by RoseTTAFold. This enhancement in predictive accuracy is further evidenced by the success rate metric (DockQ > 0.23), where HelixFold-Multimer attained a success rate of 68.9\%. In contrast, AlphaFold 3 exhibited a success rate of 59.0\% for the same threshold. The structural analysis of protein-peptide complexes profoundly influences the research and development of peptide therapeutics. Understanding the intricate interplay between proteins and peptides not only aids in the design of more effective drugs but also sheds light on fundamental biological processes, paving the way for innovative therapeutic strategies.

\subsection{Results of Antigen-Antibody Version}
Antigen-antibody structure prediction holds paramount importance in comprehending immune responses and driving advancements in drug design, vaccine development, and diagnostic assays. It stands as a fundamental pillar in the progression of biomedical research and the enhancement of therapeutic interventions.

In evaluating HelixFold-Multimer for its efficacy in predicting antigen-antibody interfaces, we embark on a comprehensive assessment. Initially, we gauge the model's performance not only on the antigen-antibody interface but also on the interface between the antibody heavy and light chains. Subsequent to this assessment, we undertake a comparative analysis to discern the impact of accurately delineating epitope regions within the specified antigen. Further analysis is undertaken on the model's confidence and efficacy across different antigen-antibody categories. Finally, we provide a case study analysis.

\subsubsection{Overall Comparison}
\begin{figure}
    \centering
    \includegraphics[width=1.0\linewidth]{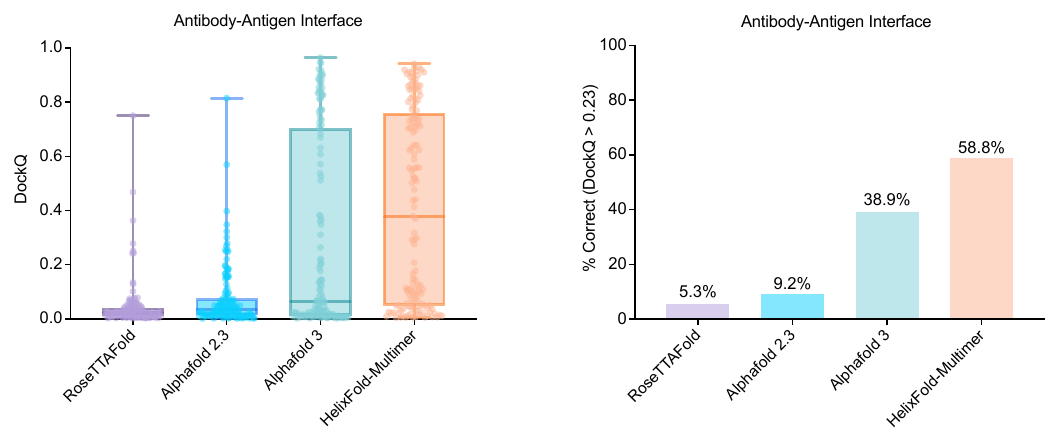} 
    \caption{Overall comparison for antibody-related complex structure predictions. The left figure depicts the distribution of DockQ scores, serving as a metric for the quality of predicted structures. The right figure showcases the percentage of accurate predictions specifically concerning antibody-antigen interfaces. The data underscores a notable improvement in prediction accuracy with the HelixFold-Multimer model compared to RoseTTAFold, AlphaFold 2.3, and AlphaFold 3.}
    \label{fig:results_dockq_abag_nano}
\end{figure}

Figure \ref{fig:results_dockq_abag_nano} illustrates the results of HelixFold-Multimer and several strong baseline methods on antibody-antigen interfaces. The median DockQ score of HelixFold-Multimer stands at 0.38, which notably outperforms AlphaFold 3 (0.07). Additionally, HelixFold-Multimer demonstrates a notable success rate of 58.8\% in accurately predicting antibody-antigen structures. This success rate surpasses that of AlphaFold 3 (38.9\%) and RoseTTAFold (5.3\%). These results indicate the superior predictive capability of HelixFold-Multimer, offering promising implications for antibody-antigen interaction studies.

\begin{figure}
    \centering
    \includegraphics[width=1.0\linewidth]{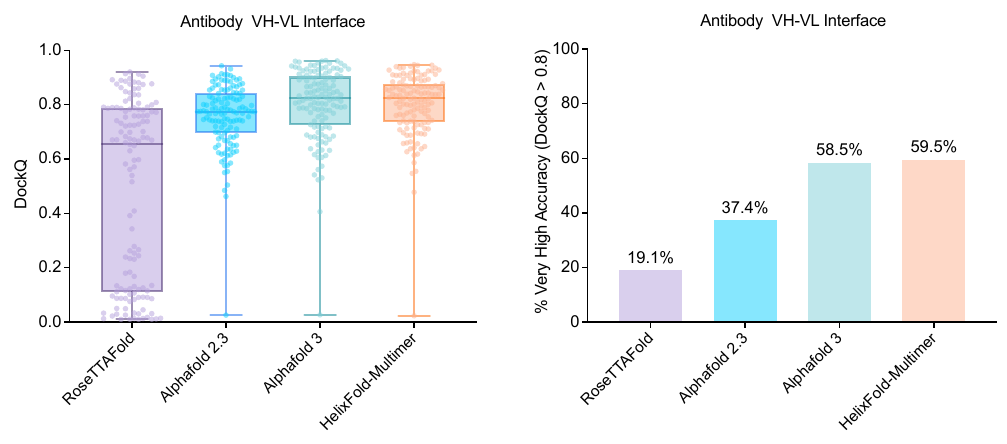} 
    \caption{Evaluation of antibody VH-VL interfaces: Box plots on the left illustrate the performance comparison among RoseTTAFold, AlphaFold 2.3,  AlphaFold 3 and HelixFold-Multimer. The bar graph on the right quantifies the percentage of predictions with very high accuracy (DockQ > 0.8).}
    \label{fig:results_dockq_ab}
\end{figure}

We subsequently evaluate the predictive accuracy for antibody VH-VL interfaces, as shown in Figure \ref{fig:results_dockq_ab}.
While existing protein structure prediction tools have achieved decent accuracy in predicting the conformation of antibodies (including heavy chain H and light chain L), higher-precision prediction results remain highly meaningful for antibody understanding. The prediction of the VH-VL chain interface by HelixFold-Multimer and AlphaFold 3 yielded the same median DockQ score of 0.823, surpassing AlphaFold 2.3's score of 0.774 and RoseTTAFold's score of 0.653. HelixFold-Multimer achieves a very high accuracy rate (DockQ > 0.8) of 59.5\%, slightly higher than Alphafold 3's 58.5\%, representing a significant improvement over AlphaFold 2.3's 37.4\% and RoseTTAFold's 19.1\%. These results indicate the proficiency of HelixFold-Multimer in accurately modeling antibody structures.

HelixFold-Multimer's outstanding ability to predict antibody-related structures suggests its potential to streamline the identification and development of new antibody-based therapeutics. This proficiency has the potential to transform the landscape of drug discovery, providing more efficient and accurate techniques for designing therapeutic antibodies customized for specific targets and diseases.

\subsubsection{Impact of Epitope Specification for Antigen-Antibody Structure Prediction}

\begin{figure}
    \centering
    \includegraphics[width=1.0\linewidth]{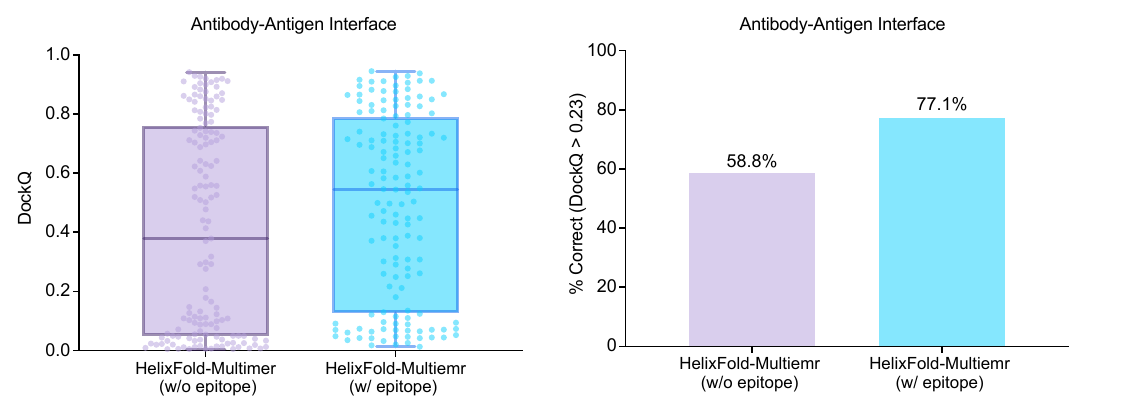} 
    \caption{Impact of epitope specification for antigen-antibody structure prediction.}
    \label{fig:results_pocket_epitope}
\end{figure}

The designation of antigen epitopes plays a crucial role in predicting antigen-antibody structures. It assists in accurately locating the binding interface, enhancing model accuracy and prediction specificity. Typically, researchers can determine antigen binding sites through analyses such as deep mutational scanning (DMS). We investigate whether there will be further improvements in accuracy after specifying the antigen epitope for HelixFold-Multimer. Figure \ref{fig:results_pocket_epitope} indicates that after specifying the antigen epitope, there is a further improvement in the prediction accuracy of HelixFold-Multimer.

\subsubsection{Model Confidence}
\begin{figure}
    \centering
    \includegraphics[width=1.0\linewidth]{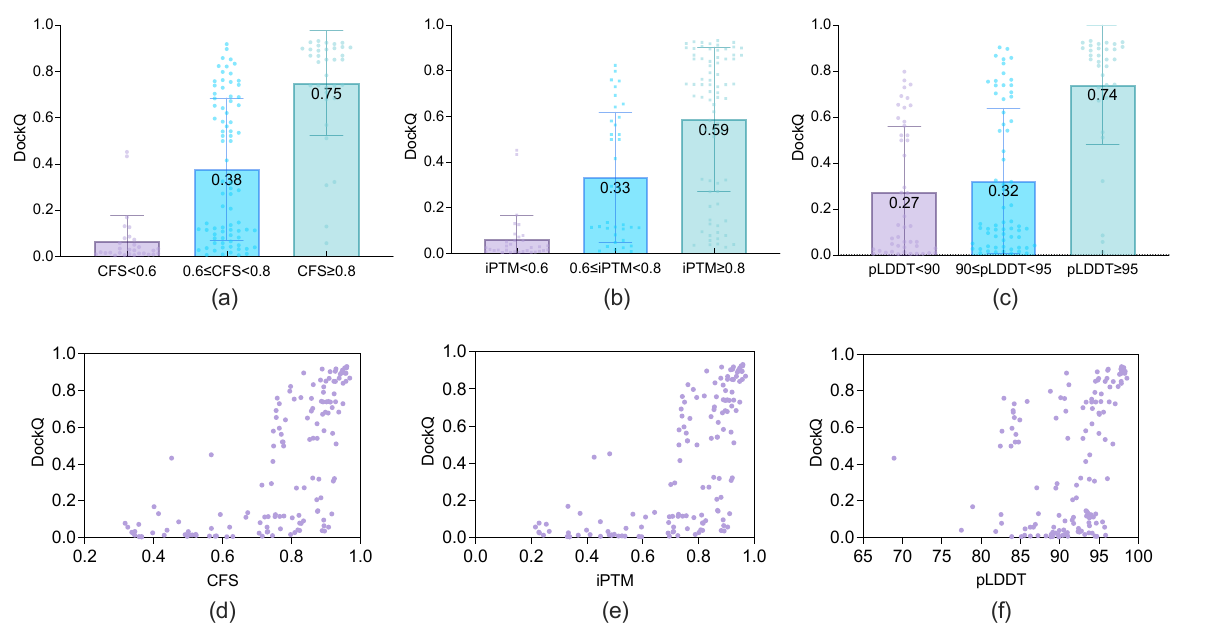} 
    \caption{Correlational analysis between the DockQ scores and the predictive indicators outputted by HelixFold-Multimer for antigen-antibody docking. Subfigures (a) to (c) present box plots illustrating the distribution of DockQ scores across different levels of predictive confidence score, iPTM, and pLDDT scores. Meanwhile, subfigures (d) to (f) depict scatter plots illustrating the associations between DockQ scores and the corresponding predictive metrics. Note that "CFS" is an abbreviation for "confidence score". 
    }
    \label{fig:results_score}
\end{figure}
The correlation between confidence scores and model accuracy is exhaustively examined to evaluate the reliability of the HelixFold-Multimer in predicting antigen-antibody interactions. 

We perform an analysis of three scoring metrics within the HelixFold-Multimer model, which assesses the confidence levels of the inference results: confidence scores (CFS), iPTM scores, and pLDDT scores. Subsequently, we organize the test samples into groups based on these confidence scores and present the distribution of DockQ scores within each group (see Figure~\ref{fig:results_score}(a)-(c)). The figures suggest a clear pattern: in general, antigen-antibody complexes identified by the model as having high confidence levels tend to have significantly higher DockQ scores compared to those with lower confidence levels.
Figure~\ref{fig:results_score}(d)-(f) elucidates the relationship between the accuracy of test samples and scoring metrics through scatter plots. All the scoring metrics show correlations with DockQ. Notably, confidence scores (Pearson correlation: 0.664) and iPTM (Pearson correlation: 0.658) exhibit stronger correlations with DockQ compared to pLDDT (Pearson correlation: 0.344). This robust correlation between scoring metrics and DockQ scores will serve as valuable guidance for leveraging HelixFold-Multimer in antibody development. For instance, analyzing these scoring metrics can help identify antibodies or antigens with higher research potential.

\subsubsection{Efficacy on Various Antigen-Antibody Categories}
\begin{figure}
    \centering
    \includegraphics[width=1.0\linewidth]{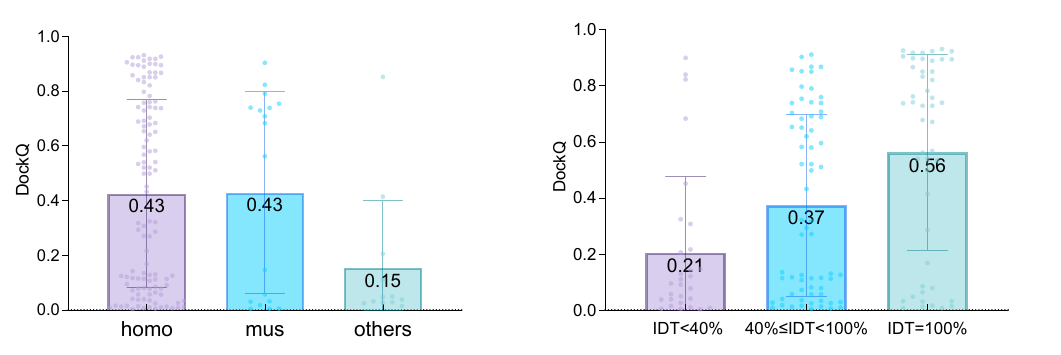} 
    \caption{Performance analysis on various antigen categories. The left figure shows the performance of antigens from different species. The right figure shows the performance of antigens with different sequence identities to the training samples. "IDT" is short for "sequence identity".
    }
    \label{fig:results_identity}
\end{figure}
We conduct an assessment of HelixFold-Multimer's effectiveness across various species groups and sequence identity intervals to identify its strengths in specific Antigen-Antibody categories. Species were classified into three groups: Homo sapiens, Mus musculus, and other species (including macaca mulatta, rattus norvegicus, synthetic construct, gallus gallus, and oryctolagus cuniculus), based on sample counts. The classification of species is based on the origin of the antibody heavy chains. The sequence identity is calculated by MMseqs2 \cite{steinegger2017mmseqs2} on antigen chains.

Different species and sequence similarities exhibit significant differences. As shown in the left subfigure of Figure \ref{fig:results_identity}, HelixFold-Multimer achieves higher mean DockQ score for Homo sapiens (0.43) and Mus musculus (0.43) compared to other species (0.15). This outcome aligns with expectations, given that the majority of antibody data originate from the research of Homo sapiens and Mus musculus. When evaluating antigens grouped by sequence similarity (the right subfigure of Figure \ref{fig:results_identity}, we observe a strong correlation between antigen sequence similarity and accuracy, as expected. HelixFold-Multimer achieves a mean DockQ score of up to 0.56 for samples corresponding to antigens already present in the training dataset. We believe that HelixFold-Multimer's performance on popular antigen-antibody complexes is already quite reliable and can effectively assist researchers in analyzing interactions between antigens and antibodies. In the future, we aim to further enhance the model's performance on less popular targets.

\subsubsection{Case Study}


We conduct visualizations using two significant viral pathogens, namely HIV and SARS-CoV-2, as benchmarks to compare the predictive capabilities of HelixFold-Multimer and AlphaFold 3. In both instances, HelixFold-Multimer exhibits exceptional performance by accurately pinpointing the binding interfaces and offering highly precise structural predictions. This precision underscores its ability to precisely capture the complex interplay between the viral antigens and antibody proteins. Conversely, AlphaFold 3's predictions fall short, as it either failed to identify the binding interfaces altogether or produced structural deviations, indicating potential limitations in its ability to accurately model antigen-antibody interactions.


\begin{figure}
    \centering
    \includegraphics[width=1.0\linewidth]{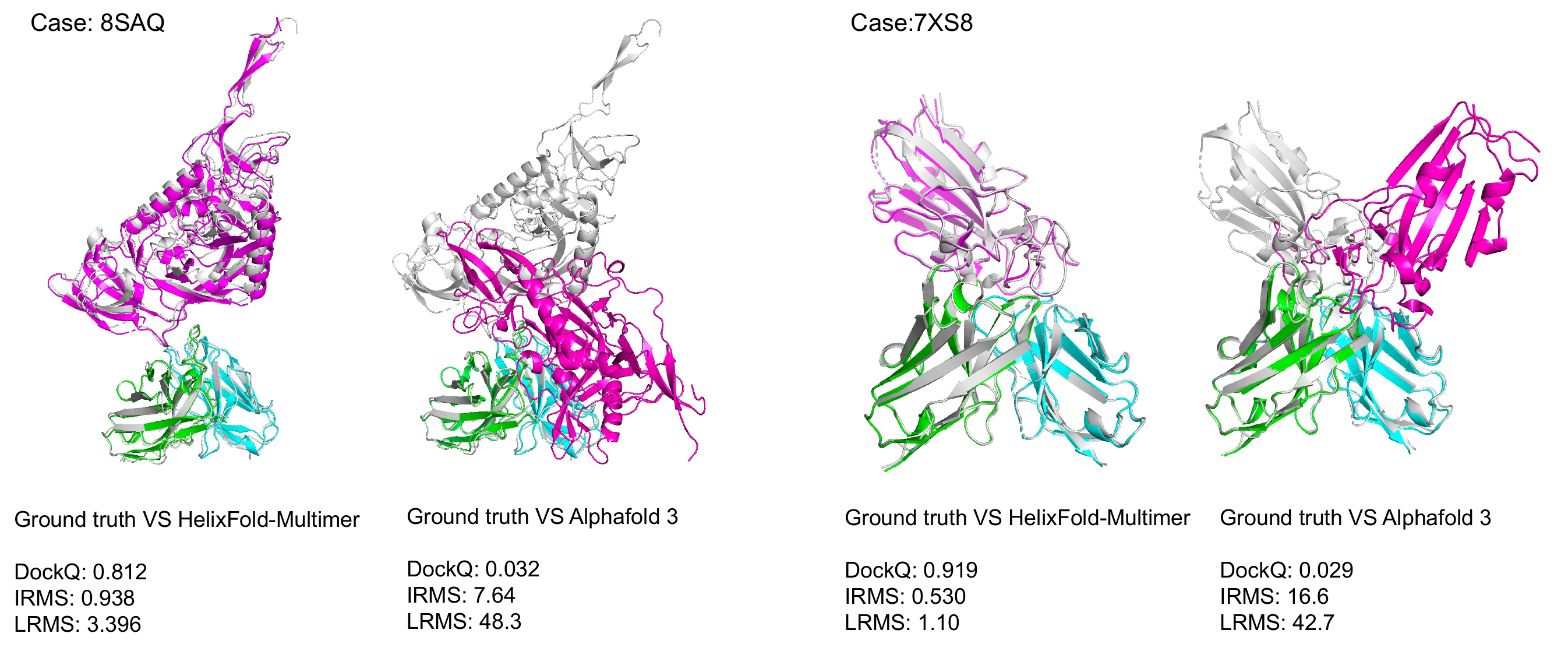} 
    \caption{Comparison of Antibody-Antigen Predictions by HelixFold-Multimer and AlphaFold 3. This figure illustrates the structural predictions for two antibody-antigen complexes: one with HIV (8SAQ) and the other with SARS-CoV-2 (7XS8). Each set shows a comparison between the actual structure (ground truth). HelixFold-Multimer demonstrating closer alignment to the true structures compared to AlphaFold 3.}
    \label{fig:results_case}
\end{figure}

\section{Conclusion}
HelixFold-Multimer showcases exceptional prowess in predicting the structures of diverse protein complexes, surpassing prior methodologies by a significant margin. Particularly noteworthy is its outstanding accuracy in forecasting the structures of protein-peptide and antigen-antibody complexes, crucial components in therapeutic interventions. These findings not only highlight the model's potential in advancing drug design for macromolecular therapeutics but also underscore its pivotal role in shaping the future landscape of therapeutic development. By harnessing the predictive power of HelixFold-Multimer, researchers are poised to unlock new possibilities in therapeutic innovation.

Currently, HelixFold-Multimer is available on the PaddleHelix platform, a comprehensive drug development service platform. Researchers can conveniently utilize the platform for inferential analysis of protein interactions. In the future, we aim to further enhance the accuracy of antibody and peptide-related structure predictions and apply them to a broader spectrum of drug development tasks. Stay tuned for more updates.

\clearpage

\begin{appendices}
\renewcommand{\thesection}{\arabic{section}}

\counterwithin{figure}{section}
\counterwithin{table}{section}
\counterwithin{equation}{section}

\section{Experimental Settings}


\subsection{Baselines}
AlphaFold and RosettaFold are currently the most representative end-to-end protein structure prediction models in the industry, serving as benchmarks. Therefore, we have chosen them as our comparison baseline.

RosettaFold \cite{humphreys2021computed} integrates deep learning with biochemical knowledge, showing promising results in single-chain protein structure prediction. However, its accuracy in predicting multi-chain protein complex structures is not satisfactory. On the other hand, AlphaFold \cite{jumper2021highly,evans2021protein}, particularly its Multimer version \cite{evans2021protein}, AlphaFold-Multimer, performs exceptionally well in predicting general protein complex structures. Nonetheless, its accuracy drops significantly in scenarios lacking sufficient cross-chain evolutionary information, such as antigen-antibody complexes. The recent released Alphafold 3\cite{abramson2024accurate} largely improved the predictive performance of antigen-antibody complex.

For the sake of convenience in comparison, this report contrasts the results of single-model inference rather than using ensemble methods (the original AlphaFold 2.3 employs inference from 25 models followed by ensemble). We conducte inference using version 2.3 model 1 from AlphaFold's GitHub repository. For AlphaFold 3, we use the official online server to generate predictions and calculate metrics based on the model 0 result that has the highest confidence score among the five predicted outcomes.
For the RoseTTAFold model, we refer to the setup described in RoseTTAFold2 \cite{baek2023efficient}.

\subsection{Evaluation Metrics}
\label{sec:Metrics}

To evaluate the accuracy of predicted protein complexes, we employ the DockQ metric \cite{basu2016dockq}. Consistent with common practice, we utilize the DockQ threshold to gauge the success rates of model predictions. Predictions with a DockQ value above 0.23 were deemed accurate, while those surpassing 0.8 were categorized as exhibiting very high accuracy.

\subsection{Confidence Metrics}
In our study, three scoring metrics are used: pLDDT (the predicted local distance difference test) \cite{mariani2013lddt}, PTM (the predicted TM-score) \cite{zhang2005tm}, and confidence score. pLDDT and PTM are metrics provided by the model and are indicators of overall structural accuracy. For interface accuracy of different chains, we measure via iPTM (the interface predicted TM-score), which is calculated based on the residues at the interface. Additionally, we use the confidence score\cite{evans2021protein} index intended to simultaneously consider the accuracy of the inter-chain interface and the overall prediction accuracy. The confidence score is calculated by assigning different weights to the PTM and iPTM values.
\subsection{Sequence Identity}
To comprehensively grasp the impact of sequence similarity on model performance, we employ sequence identity as the similarity metric. Sequence identity is defined as the percentage of residues in the evaluation set chain that match those in the training set chain. We utilize the default configuration settings of the MMSeqs2\cite{steinegger2017mmseqs2} software suite for this computation.

\section{Datasets}
\subsection{General Version}

Heteromeric protein complexes released from the period between January 16th, 2022, and December 12th, 2022, are gathered from Protein Data Bank (PDB) \cite{burley2017protein} as the evaluation set. Proteins with more than 1400 residues are excluded. These test samples are not present in the training set. The evaluation set for heteromeric proteins consists of 192 complexes.

The heteromeric protein test PDBs: 7a67  7aam  7acv  7acw  7acz  7al7  7b5g  7dz9  7dz9  7dz9  7dz9  7dz9  7dz9  7e1z  7e7c  7e9m  7eic  7eid  7eqc  7etr  7ewm  7f03  7f03  7f28  7f4n  7f4p  7f4s  7f4t  7fad  7fax  7fb8  7fba  7fc0  7fc0  7fc0  7fc0  7fc0  7fgj  7fiw  7kqk  7kql  7lo7  7mhr  7mrz  7n05  7n0a  7n0e  7nf8  7o0z  7ocj  7ocy  7od9  7oix  7oju  7oly  7ow1  7oxn  7pb4  7pb8  7pb8  7pb8    7pp2  7pqq  7pyt  7q4q  7q51  7qbe  7qbs  7qih  7qii  7qla  7qqy    7qu1  7quc  7quu  7qy5  7qy5  7r1z  7r1z  7r5a  7rc6  7rlw  7rlx  7rlz  7rm1  7rm3  7rqq  7rqr  7rsw  7ru6  7rx0  7rxq  7sch  7sd2  7sej  7sem  7sjp  7sk6  7sp8  7sp8  7srs  7ssc  7syy  7syy  7t2p  7t92  7t92  7t92  7tcq  7tcq  7tcu  7te1  7tgg  7tj4  7tr3  7u0a  7u7n  7u7n  7u8g  7ubz  7ulo  7unb  7unb  7unz  7urd  7urf  7uwj  7uym  7v4w  7vad  7vad  7vbq  7vcr  7vd7  7vgg  7vgr  7vkb  7vlz  7vmc  7wek  7wlw  7wmv  7wq3  7wq4  7wrs  7wuw  7wwq  7x2e  7x3c  7x4n  7x5c  7xjl  7y1r  7y3j  7y3j  7y6b  7y6c  7y6c  7y7m  7y8u  7ycu  7ypx  7yrh  7yx8  7yz9  7yzi  7z01  7zcl  7zk1  7zkw  7zkz  7zvi  7zvy  7zvy  7zw1  8a1e  8ap8  8ap8  8ap8  8b0u  8b38  8b38  8b38  8bd1  8d3v  8des  8dgq  8djk  8djk  8ds5  8gqe  8guo

To assess the models on protein-peptide complexes, we derive a subset of protein complexes from the heteromeric evaluation set based on the length of the shortest chain. We establish a criterion where the shortest chain's maximum length is set to 50 residues. Subsequently, we obtain a total of 61 instances of peptide-protein complexes.

The protein-peptide test PDBs: 7aam  7acv  7acw  7e7c  7e9m  7eic  7eid  7fax  7fb8  7fba  7fc0  7fc0  7fc0  7fgj  7kqk  7n05  7oju  7ow1  7oxn  7pb4  7q4q  7q51  7qbs  7qqy  7quu  7qy5  7rc6  7rlw  7rlx  7rlz  7rm1  7rm3  7rqq  7rqr  7rsw  7rxq  7sjp  7ssc  7tcq  7u0a  7ulo  7urd  7urf  7uym  7v4w  7vcr  7vd7  7vlz  7wek  7wq3  7wq4  7wwq  7x2e  7x5c  7xjl  7y3j  7y3j  7y6c  7yx8  8des  8dgq

\subsection{Antigen-Antibody Version}
We select samples from the SAbDab \cite{dunbar2014sabdab} database with release dates between January 25, 2023, and August 9, 2023, as the evaluation set for antibody-related data, ensuring that the evaluation samples are not present in our training set. Further, the antigen type of peptide and protein are chosen. Additionally, we exclud samples containing more than 1400 residues. 
In cases where AlphaFold 3 predictions failed, we intersect the successful predictions from both AlphaFold 3 and HelixFold-Multimer.
The test set comprises 141 antigen-antibody complexes. The antibody chains from antigen-antibody complexes are extracted to serve as a test dataset for the evaluation of antibody variable regions (VH-VL). We extract only the fragment variable regions of the heavy and light chains of the antibodies as inputs for the models.

The antibody-antigen test PDBs: 7u9e 8ium 8iw9 8dzv 8dy1 8iuk 8dy5 8axh 7yue 7ua2 8cz8 8db4  8hc4  8epa  8hc5  8hhy  8hhx  7uxl  7xdk  7xcz  7xda  8ek1  8eka  7xdl  8c3v  8bcz  8d7e  7wnb  7wn2  8i5i  8saw  7umn  7xj6  8dto  8sb0  8sb1  8saq  8sar  7zjl  8say  8i5h  8sb5  8sas  8sb3  8sb2  7uja  7uow  8ct6  8sav  7xj8  7xj9  8sax  7xik  8heb  7xs8  8hec  7xil  8hed  8a96  8gs9  8f0h  8fax  8fg0  8a99  7xsa  7xrz  8bse  7yru  8bsf  7zqt  8dn6  8hn6  8hn7  7st5  8dn7  8f6o  8gb8  8gb6  8ahn  8f60  8cwi  8cwj  8gb7  8dnn  8cwk  7xsc  8f6l  8av9  8g3n  8g3v  8g3r  7xeg  7xjf  8gnk  8g3q  8ee1  8ee0  8g3p  8g3o  8dwy  8g3z  8g3m  8dww  7quh  8g30  8h07  8czz  7xei  8a44  8d9y  8eoo  8da0  8da1  8d9z  8e1m  8scx  8smt  7zoz  8e1g  8duz  8elo  8elq  8de4  8cim  8elp  8gtp  7yd1  8byu  8gtq  7y8j  8el2  8ol9  7yds  8ib1  7yv1  8dz3  8dyx  7trh  8e6j  7yk4  8e6k

\end{appendices}
\clearpage

\bibliographystyle{unsrt}  
\bibliography{references}

\end{document}